\documentclass[conference,a4paper]{IEEEtran}
\usepackage{hyperref}
\usepackage{amssymb,epsfig,graphics, graphicx, url,times,mathrsfs,algorithm,algorithmic,amsmath,array,latexsym,fancyhdr,xspace,wrapfig, xcolor,multirow,amsthm,caption,cite,helvet,sidecap}

\usepackage[normalem]{ulem}

\usepackage[inline]{enumitem}
\usepackage{arydshln}
\usepackage[makeroom]{cancel}
\usepackage{varwidth}

\usepackage{tikz}
\usetikzlibrary{shapes.geometric}
\usetikzlibrary{shapes,snakes,patterns}
\usetikzlibrary{arrows,positioning,automata,calc}
\usetikzlibrary{plotmarks}

\definecolor{green1}{rgb}{0,0.5,0}
\definecolor{magenta}{rgb}{1.0, 0.11, 0.81}
\definecolor{mulberry}{rgb}{0.77, 0.29, 0.55}
\definecolor{xgray}{rgb}{0.9, 0.9, 0.9}
\def \blue{\color{blue}}
\def \red{\color{red}}

\def \Redit{\color{black}}

\def \bes{\begin{equation*}}
\def \ees{\end{equation*}}
\def \bas{\begin{align*}}
\def \eas{\end{align*}}
\def \be{\begin{equation}}
\def \ee{\end{equation}}
\def \bbm{\begin{bmatrix}}
\def \ebm{\end{bmatrix}}
\def \cf{\mathcal{E}}
\def \ck{\mathcal{R}}

\def \cd{\mathcal{D}}

\def \mrS{\texttt{S}}
\def \mrW{\texttt{I}}

\def \file {\texttt{F}}

\newtheorem{theorem}{Theorem}

\newtheorem{proposition}{Proposition}
\newtheorem{definition}{Definition}

\newtheorem{example}{Example}

\IEEEoverridecommandlockouts

\begin{document}

\title{Staircase-PIR: Universally Robust Private Information Retrieval}

\author{Rawad Bitar,~\emph{Student~Member,~IEEE} and Salim El Rouayheb,~\emph{Member,~IEEE}%
\thanks{This work was supported in part by NSF Grant CCF 1817635.
The authors are with the ECE department of Rutgers University, Piscataway, New Jersey.
Emails: firstname.lastname@rutgers.edu.}}
\vspace{-1cm}

\renewcommand{\today}{}
\maketitle
\begin{abstract}
We consider the problem of designing private information retrieval (PIR) schemes on data of $m$ files replicated on $n$ servers that can possibly collude. We focus on devising robust PIR schemes that can tolerate stragglers, i.e., slow or unresponsive servers. 
In many settings, the number of stragglers is not known a priori or may change with time. We define universally robust PIR as schemes that achieve PIR capacity asymptotically in $m$ and simultaneously for any number of stragglers up to a given threshold. We introduce Staircase-PIR schemes and prove that they are universally robust. Towards that end, we establish an equivalence between robust PIR and communication efficient secret sharing.

\end{abstract}

\section{Introduction}
We consider the problem of designing PIR schemes on replicated data \cite{PIR1995, chor1998private}, i.e., the data consisting of $m$ files is {\em replicated} on $n$ servers that can possibly collude. A user queries the servers to obtain a file of interest while keeping private the identity of the file, even if $t$ servers collude with $t<n$. We focus on information-theoretic privacy (instead of computational privacy), which guarantees privacy without making any assumption on the computation power of the servers as long as no more than $t$ servers collude.

\noindent{\em Robustness:} Under the setting described above, we are interested in PIR schemes that are robust. A robust PIR scheme allows the user to retrieve the file by receiving responses from any $k$ servers $t<k\leq n$. The robustness property is motivated by the need to mitigate the effect of stragglers \cite{dean2008mapreduce,DB13,TOFEC,JLS12,KSS15,speeding,BPR18,tandon2016gradient,yu2017polynomial,halbawi2017improving,dutta2017coded,li2018polynomially}.

\noindent{\em PIR capacity:} A challenge in designing PIR schemes is maximizing their download rate defined as the ratio of the retrieved file size to the amount of information downloaded by the user. The PIR capacity is defined as the maximal rate achievable by a PIR scheme and was shown in \cite{SJ18} to be
\begin{equation}\label{eq:cap}
C_m(t,k) = \frac{1-t/k}{1-\left(t/k\right)^m}.
\end{equation}
Note that $C_m(t,k)$ converges exponentially with the number of files $m$ to its asymptotic value
\begin{equation}\label{eq:acap}
C(t,k) = 1-\dfrac{t}{k}.
\end{equation}
For instance, for $m=3$ files and $k=10$ servers, $C_3(1,10)$ is at $99\%$ 
of its asymptotic value $C(1,10)$ (assuming no collusion $t=1$). Given the tremendous amount of files being stored in current systems, we focus on PIR schemes achieving the asymptotic capacity given in~\eqref{eq:acap}.

\noindent{\em Universality:} In many cases, the exact number of stragglers is not known a priori. This motivates the study of universally robust PIR schemes that can tolerate a varying number of stragglers. More precisely, a universally robust PIR scheme allows the user to query $n$ servers and retrieve the wanted file from any $\mu$ responses, $k\leq \mu \leq n$, while achieving the asymptotic PIR capacity $C(t,\mu)$ given in~\eqref{eq:acap} simultaneously for all possible values of $\mu$. Although a robust PIR scheme designed for the worst case number of stragglers ($\mu=k$) can tolerate up to $n-k$ stragglers, it does not necessarily achieve the capacity for all possible values of $\mu$.

In the case where the stragglers are non-responsive, universally robust PIR guarantees that the user can always obtain the file even if up to $n-k$ servers do not reply to its queries. On the other hand, when the stragglers are slow servers, universally robust PIR offers the user a tradeoff between download rate and waiting time. The user can pick the number of servers to wait for and achieve asymptotic PIR capacity corresponding to this choice. We illustrate the idea in Example~\ref{ex:uni}.

\begin{table*}[t!]
\centering
\begin{tikzpicture}
\definecolor{ceruleanblue}{rgb}{0.16, 0.32, 0.75}
\definecolor{darkcandyapplered}{rgb}{0.64, 0.0, 0.0}
\definecolor{ogreen}{rgb}{0,0.5,0}
\def\ogreen{\color{ogreen} \bf}

\node[font = \small] (t) at (0,0) {\centering
\renewcommand{\arraystretch}{1.2}
\begin{tabular}[h!]{c|c|c|c}
~& Server1 & Server 2 & Server 3\\ \hline
Storage & $\mathbf{x}$ & $\mathbf{x}$ & $\mathbf{x}$ \\ \hline
\multirow{2}{*}{Queries}&   $\mathbf{r}_1$ &  $\mathbf{e}_{i}+\mathbf{r}_1$ &  $2\mathbf{e}_{i}+\mathbf{e}_{m+i}+\mathbf{r}_1$\\
~& $\mathbf{r}_2$ &  $\mathbf{e}_{m+i}+\mathbf{r}_2$ & $2\mathbf{e}_{m+i}+\mathbf{r}_2$ \\ \hline
\multirow{2}{*}{Responses} & \blue $\mathbf{r}_1^T\mathbf{x}$ & \blue $(\mathbf{e}_{i}+\mathbf{r}_1)^T\mathbf{x}$ & \blue $(2\mathbf{e}_{i}+\mathbf{e}_{m+i}+\mathbf{r}_1)^T\mathbf{x}$\\
~& $\mathbf{r}_2^T\mathbf{x}$ &  $(\mathbf{e}_{m+i}+\mathbf{r}_2)^T\mathbf{x}$ & $(2\mathbf{e}_{m+i}+\mathbf{r}_2)^T\mathbf{x}$ \\
\end{tabular}
};

\draw[dashed] (5,2) -- (5,-2);

\node[above right =-1 and 0.5 of t] (x) {$\mathbf{x}=\begin{bmatrix} x_1,\dots,x_m, x_{m+1},\dots,x_{2m} \end{bmatrix}^T$};	
\node[below=0.2 of x] (f) {$f_i = [x_i,\ x_{m+i}]=[\mathbf{e}_i^T\mathbf{x},\ \mathbf{e}^T_{m+i}\mathbf{x}]$};
\node[below = 0.2 of f] (e) {$\mathbf{e}_i= [{0},\dots,0,\underbrace{1}_{\textstyle i^\text{th}},0,\dots,{0}]^T$};
\node[below right=-0.4 and -1.6 of e] {entry};
\node[below = 0 of e] (r) {$\mathbf{r}_1$ and $\mathbf{r}_2$: random vectors};
\node[above=0.3 of x] (n) {\em \large Notation};

\end{tikzpicture}
\vspace{0.1cm}
\caption{An example of Staircase-PIR code for $n=3$, $k=2$ and $t=1$. 
The user sends $2$ sub-queries to each server. Each server projects the data on the sub-query vectors and sends the result to the user. If all $3$ servers are not stragglers, the user only downloads the first response (in blue) from each server to retrieve the required file as follows $x_{i} = (\mathbf{e}_{i}+\mathbf{r}_1)^T\mathbf{x} - \mathbf{r}_1^T\mathbf{x}$ and $x_{m+i} =(2\mathbf{e}_{i}+\mathbf{e}_{m+i}+\mathbf{r}_1)^T\mathbf{x} - \mathbf{r}_1^T\mathbf{x}-2x_{i}$. The rate of this scheme is equal to $C(1,3)=2/3$, because the user downloads $3$ responses to retrieve the $2$ parts of the file $f_i=[x_i,\ x_{m+i}]$. If $1$ server is a straggler, the user downloads $2$ responses from each of the remaining servers to retrieve the file. The user can retrieve the file irrespective of which server is a straggler. The rate of this scheme is $C(1,2)=1/2$, because the user downloads $4$ responses to retrieve the file.}
\label{fig:intro}
\vspace{-0.5cm}
\end{table*}

\begin{example}[Universally robust Staircase-PIR]
\label{ex:uni} %
Let $n=3$ servers and we want to tolerate up to $1$ straggler, i.e., $k=2$. Assume the servers do not collude, i.e., $t=1$. We describe the Staircase-PIR that simultaneously achieves $C(1,2)=1/2$ in the case of $1$ straggler, $\mu=2$, and $C(1,3)=2/3$ in the case of no stragglers, $\mu=3$.

 Let $f_1,\dots,f_m$ be the files stored on the servers. We divide each file $f_i$ into 2 parts and denote the $i^\text{th}$ file as $f_i = [x_i,\ x_{m+i}]$. We represent the data by the vector\footnote{All vectors are column vectors and the superscript $T$ denotes the transpose operator.} $\mathbf{x}=[x_1,\dots,x_m,x_{m+1},\dots,x_{2m}]^T$. Let $\mathbf{e}_i$ denote the all zero vector of length $2m$ with a `$1$' in the $i^\text{th}$ entry. The $i^\text{th}$ file can be expressed as \mbox{$f_i = [\mathbf{e}_i^T\mathbf{x},\ \mathbf{e}^T_{m+i}\mathbf{x}]$} which is the projection of $\mathbf{x}$ on $\mathbf{e}_i$ and $\mathbf{e}_{m+i}$. To construct the Staircase-PIR scheme, we use two independent random vectors $\mathbf{r}_1$ and $\mathbf{r}_2$ each of length $2m$ with entries drawn uniformly at random from $GF(2)$. We encode the queries using Staircase codes as shown in Table~\ref{fig:intro}.

The user sends $2$ sub-queries to each server which projects the data on the queries and sends the result to the user. If one server is straggler, the user downloads all the sub-queries from the other $2$ servers to retrieve the file. For instance, if server $3$ is the straggler, the user downloads all the responses from servers $1$ and $2$, i.e., $4$ responses in total, and retrieves $f_i=[x_i,\ x_{m+i}]$ as $x_i = (\mathbf{e}_{i}+\mathbf{r}_1)^T\mathbf{x} - \mathbf{r}_1^T\mathbf{x}$ and  $x_{m+i} = (\mathbf{e}_{m+i}+\mathbf{r}_2)^T\mathbf{x} - \mathbf{r}_2^T\mathbf{x}$. The rate in this case is equal to $2/4=1/2=C(1,2)$. However, if no server is straggler, the user only downloads the first response (in blue) of all $3$ servers to retrieve the file. The rate in this case is equal to $2/3=C(1,3)$. Privacy is achieved because $\mathbf{e}_i$ and $\mathbf{e}_{m+i}$ are padded with random vectors.
\end{example}

\noindent{\em{Related work:}} 
Private information retrieval was introduced by Chor et al. \cite{PIR1995,chor1998private} and was followed up by a large body of work, e.g., \cite{WY05,gasarch2004survey ,yekhanin2010private, beimel2002breaking,shah2014one,chan2014private,tajeddine2016private,freij2016private,kumar2017private,banawan2016capacity,SJ18,sun2016capacity}. The literature mainly focused on reducing the communication cost of privately retrieving the file. The early body of work measured the communication cost by the amount of information uploaded (queries) and downloaded (responses) by the user \cite{gasarch2004survey,WY05, yekhanin2010private, beimel2002breaking}. Given the increasing size of stored files, the recent body of work measures the communication cost by the amount of information downloaded by the user, assuming the queries are too small compared to the downloaded files \cite{shah2014one,chan2014private,tajeddine2016private,freij2016private,kumar2017private,banawan2016capacity,SJ18,sun2016capacity} which is the assumption we adopt in this paper.

Robust PIR was studied in the literature, e.g., \cite{beimel2002robust,SJ18,sun2016capacity,devet2012optimally,TR17,tajeddine2018robust,banawan2016capacity,banawan2017capacity} and the capacity of robust PIR schemes under download cost was characterized in \cite{SJ18,sun2016capacity,banawan2016capacity} for replicated data. The common focus of the literature has been on designing robust PIR that are not necessarily universal, which are tailored to a specific number of stragglers. In \cite{TR17} the authors present a universally robust PIR scheme for the no collusion case and when the data is stored on the servers using a maximum distance separable (MDS) code.

\noindent{\em Contributions:}  We introduce Staircase-PIR, a universally robust PIR scheme achieving asymptotic capacity for any number of stragglers up to a given threshold. Compared to the previous work on universal PIR \cite{TR17}, this work allows servers' collusion but is restricted to the case of replicated data. 
The main ingredient of the proposed scheme is Staircase secret sharing codes introduced by the authors in \cite{BRIT18, BR16}. Moreover, we establish an equivalence between robust PIR schemes achieving asymptotic capacity and communication efficient secret sharing schemes.

\section{Problem formulation and main results}\label{sec:sys}
We consider robust private information retrieval. The data $\mathbf{x}$ is formed of $m$ files $f_1, \dots, f_m$ and is replicated on $n$ servers. A user wants to retrieve a file $f_i$ from the data without revealing the identity $i$ of the file to the servers. 
A robust PIR scheme encodes a set of queries $\mathbf{q}_1,\dots, \mathbf{q}_n$ to be sent by the user to the servers. Let $\mrW$ denote the random variable representing the identity of the file that the user wants and let $\texttt{F}$ be the random variable representing the file $f_i$. Let $\texttt{Q}_i$ denote the random variable representing query $\mathbf{q}_i$ and let $[n]\triangleq\{1,\dots,n\}$. For any subset $B\subseteq[n]$ denote by $\texttt{Q}_B$ the set of random variables representing the queries indexed by $B$, i.e., $\texttt{Q}_B=\{\texttt{Q}_i \! : i\in B\}$. Let $H(.)$ denote the entropy function. Then, a universally robust PIR scheme is defined as follows.
\begin{definition}[Universally robust PIR]
A universally robust PIR (UR-PIR) scheme is a scheme that satisfies the following properties:

\noindent {\em 1) Privacy:} Any subset of $t$ or less queries do not reveal any information about the identity of the file, i.e., 
\begin{equation}\label{eq:secrecy}
H(\mrW \mid \texttt{Q}_T)=H(\mrW),\, \forall \ T\subset [n] \text{ s.t.}   \left\lvert T \right\rvert = t.
\end{equation}

\noindent {\em 2) Robustness:} When receiving the responses of any $\mu$ servers, $ k\leq \mu \leq n$, the user obtains the file,
\be 
\label{eq:mds}
H(\file \mid \texttt{Q}_A)=0,\, \forall A\subseteq [n] \text{ s.t. }  k\leq \left\lvert A\right\rvert \leq n.
\ee

\noindent {\em 3) Optimality:} The capacity of robust PIR is \cite{SJ18}
\begin{align}
C(t,k)=1-\frac{t}{k}.
\label{eq:acap1}
\end{align}
We say that the scheme is optimal if the rate of the scheme achieves $C(t,k)$ given in~\eqref{eq:acap1}, i.e.,
\be
\label{eq:CO}
\dfrac{H(\file)}{H(\texttt{Q}_A)}=1-\dfrac{t}{\mu}, \, \forall A \subset [n] \text{ s.t. }  \left\lvert A \right\rvert =\mu,\ee
for all $\mu$, $k\leq \mu \leq n$.
\end{definition}

In addition, we refer to a robust PIR achieving capacity $C(t,k)$ as an $(n,k,t)$ robust PIR and we refer to a universally robust PIR achieving capacity $C(t,\mu)$ for all $k\leq \mu \leq n$ as an $(n,k,t)$ UR-PIR. We introduce Staircase-PIR, a deterministic construction for all $(n,k,t)$ UR-PIR schemes.
\begin{theorem}\label{thm:main}
The $(n,k,t)$ Staircase-PIR scheme described in Section~\ref{sec:staircase} is a universally robust PIR, i.e., satisfies the required privacy and robustness constraints given in~\eqref{eq:secrecy}~and~\eqref{eq:mds} {\Redit for any given $t< k \leq n$}, and achieves the asymptotic capacity of robust PIR
\begin{align*}
C(t,\mu)=1-\frac{t}{\mu},
\end{align*}
simultaneously for all $\mu$ such that $k \leq \mu \leq n$.
\end{theorem}


\section{Staircase-PIR scheme} \label{sec:cons}
\subsection{Staircase-PIR construction} \label{sec:staircase}
We describe the $(n,k,t)$ Staircase-PIR scheme. The scheme consists of three steps: \begin{enumerate*} \item the user encodes the queries $\mathbf{q}_1,\dots,\mathbf{q}_n$ and sends them to the servers; \item each server $i$, $i=1,\dots,n$, projects the data on the received query, i.e., computes $\mathbf{q}_i^T\mathbf{x}$ and sends the result to the user; and \item the user decodes the requested file\end{enumerate*}. We start by explaining the encoding of the queries. Let $\mu_j=n-j+1$, and  
 $\alpha_j=\mu_j-t,$ $j=1,\dots, n-k+1$. Staircase-PIR divides each query into $\alpha$ sub-queries, where $\alpha=LCM(\alpha_1,\dots,\alpha_{n-k})$ is the least common multiple of all the $\alpha_j$'s except the last $\alpha_{n-k+1}=k-t$. Consequently, the construction assumes that each file $f_i$ of the data is divided into $\alpha'=(k-t)\alpha$ parts, i.e., $f_i=[x_i,x_{m+i},\dots,x_{(\alpha'-1)m+i}]= [\mathbf{e}_i^T\mathbf{x},\dots,\mathbf{e}_{(\alpha'-1)m+i}^T\mathbf{x}]$ for $i=1,\dots,m,$ and \mbox{$\mathbf{x}=[x_1,\dots,x_m,x_{m+1},\dots,x_{\alpha'm}]$}. Let $\mathbf{e}'_{j,i}\triangleq \mathbf{e}_{(j-1)m+i}$, we drop the index of the requested file $i$ when it is clear from the context.

\noindent To retrieve file $i$, the user encodes $\mathbf{e}'_1,\dots,\mathbf{e}'_{\alpha'}$ together with $t\alpha$ iid random vectors $\mathbf{r_1},\dots,\mathbf{r}_{\alpha'}$ into $n$ queries. The random vectors are drawn uniformly at random from $GF(2)$ and independently of the $\mathbf{e}'_j$'s. To encode the queries, we arrange the $\mathbf{e}'_j$'s in an $\alpha_1\times \alpha'/\alpha_1$ matrix $\cf$ and the random vectors in $n-k+1$ matrices $\ck_j, j=1,\dots,n-k+1,$ of respective dimensions $t\times \alpha' / \alpha_j\alpha_{j-1}$ (take $\alpha_0=1$). Note that each entry of the matrices $\cf$ and the $\ck_j$'s is a vector of length $\alpha m$.

We arrange $\cf$ and the $\ck_j$'s in $n-k+1$ matrices $\mathcal{M}_j,$ $j=1,\dots,n-k+1,$ as follows,
\bes
\label{eq:matcon}
  \begin{tikzpicture}[baseline=(current  bounding  box.center)]
\tikzstyle{stealth} = [draw=none,text=black]
\node[stealth] (1) at (0,0){$
\mathcal{M}_1=\hspace{0.3cm} \bbm
\multirow{2}{*}{$\cf$} \\
\\
 \ck_1\\
\ebm$                 };

\node[stealth] (2) [right=0.1cm of 1] {and      $\mathcal{M}_j=\hspace{0.3cm}\bbm
\cd_{j-1} \\ \ck_{j} \\ \mathbf{0} \ebm \quad j\neq 1$,} ;
\end{tikzpicture}
\vspace{-0.2cm}
\ees
where, $\cd_l$ is a matrix of dimensions $\alpha_{l+1} \times \alpha'/\alpha_l \alpha_{l+1}$, $l=1,\dots, n-k$, formed of the $\left(n-l+1\right)^{th}$ row of $\bbm \mathcal{M}_1 \ \mathcal{M}_2 \ \cdots \ \mathcal{M}_l \ebm$ wrapped around to fit the above mentioned dimensions. Each matrix $\mathcal{M}_j$ is completed to $n$ rows with the all zero matrix $\mathbf{0}$. We obtain the encoding matrix $\mathcal{M}$ defined in Table~\ref{tab:unim} by concatenating the $n-k+1$ matrices $\mathcal{M}_j$, $j=1,\dots,n-k+1$. Note that the positions of $\ck_j$ and $\mathcal{D}_{j-1}$ (similarly $\ck_1$ and $\mathcal{E}$) in $\mathcal{M}$ can be switched without affecting the construction.

\begin{table}[h!]
\centering
\resizebox{!}{0.08\textwidth}{
\begin{tikzpicture}[baseline=(current  bounding  box.center)]
\tikzstyle{stealth} = [draw=none,text=black]


\node[stealth] at (0,0)
{$\mathcal{M}=\left[\!
\begin{array}{c:c:c:c:c} 
 ~ & ~ &\multirow{2}{*}{$\cd_2$}  & \dots & \cd_{h-1}\\
 ~& \multirow{2}{*}{$\cd_1$} & & & \multirow{2}{*}{$\ck_h$} \\
\multirow{2}{*}{$\mathcal{E}$} & & \multirow{2}{*}{$\ck_3$}  &  \dots  & \\ 
 & \multirow{2}{*}{$\ck_2$} & & & \multirow{3}{*}{$\mathbf{0}$} \\
\multirow{2}{*}{$\ck_1$} &  & \multirow{2}{*}{$\mathbf{0}$} & \dots & \\
 & \mathbf{0} & & & \\
 \end{array}\!\right].$};

 \end{tikzpicture}
 }
 \caption{ The structure of the matrix $\mathcal{M}$ used to encode the queries of Staircase-PIR. }
 \label{tab:unim}
\end{table}


\noindent{\em Encoding of the queries: } Let $\mathcal{V}$ be an $n\times n$ Vandermonde\footnote{We require all square sub-matrices formed by consecutive columns of $\mathcal{V}$ to be invertible. Two family of matrices satisfying this property are Vandermonde and Cauchy matrices.} matrix over $GF(q)$, $q>n$. The matrix $\mathcal{M}$ is multiplied by $\mathcal{V}$ to obtain the query matrix  $\mathcal{Q}=\mathcal{V}\mathcal{M}$. The queries sent to the servers are the $n$ rows of $\mathcal{Q}$. Note that each row of $\mathcal{Q}$ consists of $\alpha$ entries, hence each query is divided in $\alpha$ sub-queries.

\noindent{\em Retrieving the file:} To retrieve the wanted file by waiting for any $\mu_j$ servers indexed by $L\subseteq [n]$, the user only downloads the projection of $\mathbf{x}$ on the first $\alpha'/\alpha_j$ sub-queries from each contacted servers corresponding to $(\mathbf{v}_l \bbm \mathcal{M}_1 \ \cdots\ \mathcal{M}_j \ebm)^T\mathbf{x}$, for all $l\in L$, where $\mathbf{v}_l$ denote the $l^\text{th}$ row of $\mathcal{V}$. Decoding all the $\mathbf{e}'$'s from the received parts of the responses is guaranteed by \cite[Theorem~2]{BRIT18}, therefore retrieving the file $[\mathbf{e'}_1,\dots,\mathbf{e}'_{\alpha'}]^T\mathbf{x}$ follows from the linearity of the scheme.

\noindent{\em Optimality:} When waiting for $\mu_j$ servers, the user downloads $\mu_j \alpha'/\alpha_j$ responses to retrieve the $\alpha'$ parts of the file. Therefore, the rate of the scheme is equal to $\alpha'/(\mu_j\alpha'/\alpha_j) = \alpha_j/\mu_j = (\mu_j-t)/\mu_j = C(t,\mu_j)$ given in~\eqref{eq:acap}. 

\noindent{\em Privacy:} Each subset of at most $t$ servers obtain no information about the identity of the wanted file. Each server $i$, $i=1,\dots,n,$ observes only the query $\mathbf{q}_i$. Privacy follows from \cite[Theorem~2]{BRIT18} which guarantees that any collection of $t$ queries leak no information about the $\mathbf{e}'_j$'s.

\vspace{-0.1cm}
\subsection{Examples of Staircase-PIR}\label{sec:uniex}
First we show how the scheme in Example~\ref{ex:uni} was obtained using the general construction.

\addtocounter{example}{-1}
\begin{example}[Continued]
Recall that we want to construct an $(n,k,t)=(3,2,1)$ Staircase-PIR scheme. Each file is divided into two parts $f_i = [\mathbf{e}_i^T\mathbf{x},\ \mathbf{e}_{m+i}^T \mathbf{x}]$ and the construction uses two random vectors $\mathbf{r}_1$ and $\mathbf{r}_2$. The matrix $\mathcal{M}$ is created by arranging the vectors $\mathbf{e}_i$ and $\mathbf{e}_{m+i}$ in $\mathcal{E} = [\mathbf{e}_i,\ \mathbf{e}_{m+i}]$ and by having $\ck_1 = [\mathbf{r}_1]$, $\ck_2 = [\mathbf{r}_2]$ and $\mathcal{D}_1 = [\mathbf{e}_{m+i}]$. The matrix $\mathcal{M}$ in this example is given by
\begin{equation*}
\mathcal{M}=\bbm \mathcal{R}_1 & \mathcal{R}_2 \\
\multirow{2}{*}{$\mathcal{E}$} & \mathcal{D}_1\\
~ & 0 \ebm  =  \bbm \mathbf{r}_1 & \mathbf{r}_2 \\ \mathbf{e}_{i} & \mathbf{e}_{m+i} \\ \mathbf{e}_{m+i} & \mathbf{0} \ebm.
\end{equation*}
The user constructs the query matrix $\mathcal{Q}=\mathcal{V}\mathcal{M}$ as given in~\eqref{eq:conss}, where each row $\mathbf{q}_i$ of $\mathcal{Q}$ is the query sent to server $i$. The queries are given in Table~\ref{fig:intro}.
\begin{equation}
\label{eq:conss}
\mathcal{Q} = \mathcal{V} \mathcal{M} = \bbm 1 & 0 & 0 \\ 1 & 1 & 0 \\ 1 & 2 & 1\ebm \bbm \mathbf{r}_1 & \mathbf{r}_2 \\ \mathbf{e}_{i} & \mathbf{e}_{m+i} \\ \mathbf{e}_{m+i} & \mathbf{0} \ebm.
\end{equation}
\end{example}

Next we give a second example that illustrates in more details the general construction of Staircase-PIR.

\begin{example}
\label{ex:stair}
We construct an $(n,k,t)=(4,2,1)$ Staircase-PIR. Let $\mu_1=4,\ \mu_2=3$, $\mu_3=2$, $\alpha_1=3$, $\alpha_2=2$, $\alpha_3=1$ and  $\alpha= LCM(\alpha_1,\alpha_2)=LCM(3,2)=6$. We divide the files into $\alpha'=6$ parts each, i.e., $f_i = [\mathbf{e}_i^T \mathbf{x}, \dots, \mathbf{e}_{5m+i}^T\mathbf{x}] \triangleq [{\mathbf{e}'_1}^T\mathbf{x},\dots,{\mathbf{e}'_{6}}^T\mathbf{x}]$. The construction uses $t\alpha=6$ iid random vectors $\mathbf{r}_1,\dots,\mathbf{r}_6$ drawn uniformly at random from $GF(2)$. The $\mathbf{e}'_j$'s and the random vectors are arranged in the following matrices,
$$
\cf=\bbm
\mathbf{e}'_1 & \mathbf{e}'_4\\
\mathbf{e}'_2 & \mathbf{e}'_5\\
\mathbf{e}'_3 & \mathbf{e}'_6\\
\ebm,\ \ck_1=\bbm
\mathbf{r}_1 & \mathbf{r}_2  \ebm,$$
$$\ck_2=
\bbm \mathbf{r}_3\ebm,\ 
\text{and }  \ck_3=\bbm \mathbf{r}_4&\mathbf{r}_5&\mathbf{r}_6\ebm.
$$
To build the matrix $\mathcal{M}$ which will be used for encoding the queries, we start with $$
\mathcal{M}_1=\bbm
\mathbf{e}'_1 & \mathbf{e}'_2 & \mathbf{e}'_3 & \mathbf{r}_1\\
\mathbf{e}'_4 & \mathbf{e}'_5 & \mathbf{e}'_6 & \mathbf{r}_2\\
\ebm^T.
$$
Then, $\cd_1$ is the $\alpha_2 \times \alpha'/\alpha_1\alpha_2=2\times1$ matrix containing the entries of the $n^{th}$ row of $\mathcal{M}_1$, i.e., $\cd_1=\bbm
\mathbf{r}_1& \mathbf{r}_2\ebm^T$. Therefore, $\mathcal{M}_2=\bbm \cd_1& \ck_2&\mathbf{0}\ebm^T=\bbm
\mathbf{r}_1&\mathbf{r}_2&\mathbf{r}_3&0\ebm^T$. Similarly, we have $\cd_2=\bbm \mathbf{e}'_3&\mathbf{e}'_6&\mathbf{r}_3\ebm$ and $$\mathcal{M}_3=\bbm \mathbf{e}'_3&\mathbf{e}'_6&\mathbf{r}_3 \\ \mathbf{r}_4&\mathbf{r}_5&\mathbf{r}_6\\ 0&0&0\\ 0&0&0\ebm.$$ We obtain $\mathcal{M}$ by concatenating $\mathcal{M}_1,\ \mathcal{M}_2$ and $\mathcal{M}_3$,

\begin{equation*} \label{eq:unmat}
\begin{tikzpicture}[baseline=(current  bounding  box.center)]
\tikzstyle{stealth} = [draw=none,text=black]


\node[stealth] (1) at (0,0) {$\mathcal{M}=\left[\!
\begin{array}[h!]{cccccc}
\mathbf{e}'_1 & \mathbf{e}'_4 &\red \mathbf{r}_1 & \blue \mathbf{e}'_3 &  \blue \mathbf{e}'_6 & \blue  \mathbf{r}_3 \\
\mathbf{e}'_2 & \mathbf{e}'_5 & \red \mathbf{r}_2 & \mathbf{r}_4 &  \mathbf{r}_5 &   \mathbf{r}_6 \\
\blue \mathbf{e}'_3 &\blue  \mathbf{e}'_6 & \blue  \mathbf{r}_3 &  0 &   0 &   0\\
\red \mathbf{r}_1 &\red \mathbf{r}_2 & 0 &   0 &   0 &   0 \\ 
\end{array}\!\right].$};

%
%
\end{tikzpicture}
\vspace{-0.1cm}
\end{equation*}
The matrix $\mathcal{V}$ is the $n\times n=4\times 4$ Vandermonde matrix over $GF(5)$ given in~\eqref{eq:unv}. The query $\mathbf{q}_j$ sent to server $j$ is the $j^\text{th}$ row of the matrix $\mathcal{Q}=\mathcal{V}\mathcal{M}$. The responses to the queries, i.e., $\mathcal{Q}^T\mathbf{x}$ are given in Table~\ref{tab:uniex}.

\begin{equation}
\label{eq:unv}
\mathcal{V}= \begin{bmatrix}
1 & 1 & 1 & 1\\
1 & 2 & 4 & 3\\
1 & 3 & 4 & 2\\
1 & 4 & 1 & 4\\
\end{bmatrix}.
\end{equation}
\begin{table}[t]
\vspace{-0.3cm}
\centering
\begin{tabular}[h!]{c|c}
Server 1 & Server 2 \\  \hline
$(\mathbf{e}'_1+\mathbf{e}'_2+\mathbf{e}'_3+\mathbf{r}_1)^T\mathbf{x}$ & $(\mathbf{e}'_{1}+2\mathbf{e}'_{2}+4\mathbf{e}'_{3}+3\mathbf{r}_1)^T\mathbf{x}$ \\ 
$(\mathbf{e}'_4+\mathbf{e}'_5+\mathbf{e}'_6+\mathbf{r}_2)^T\mathbf{x}$ & $(\mathbf{e}'_{4}+2\mathbf{e}'_{5}+4\mathbf{e}'_{6}+3\mathbf{r}_2)^T\mathbf{x}$ \\ 
\red $(\mathbf{r}_1+\mathbf{r}_2+\mathbf{r}_3)^T\mathbf{x}$ &\red  $(\mathbf{r}_1+2\mathbf{r}_2+4\mathbf{r}_3)^T\mathbf{x}$ \\ 
\blue $(\mathbf{e}'_{3}+\mathbf{r}_4)^T\mathbf{x}$ & \blue $(\mathbf{e}'_{3}+2\mathbf{r}_4)^T\mathbf{x}$ \\ 
\blue $(\mathbf{e}'_{6}+\mathbf{r}_5)^T\mathbf{x}$ & \blue $(\mathbf{e}'_{6}+2\mathbf{r}_5)^T\mathbf{x}$ \\ 
\blue $(\mathbf{r}_3+\mathbf{r}_6)^T\mathbf{x}$ & \blue $(\mathbf{r}_3+2\mathbf{r}_6)^T\mathbf{x}$ \\ 
\multicolumn{2}{c}{~}\\
Server 3 & Server 4\\ \hline
$(\mathbf{e}'_{1}+3\mathbf{e}'_{2}+4\mathbf{e}'_{3}+2\mathbf{r}_1)^T\mathbf{x}$&$(\mathbf{e}'_{1}+4\mathbf{e}'_{2}+\mathbf{e}'_{3}+4\mathbf{r}_1)^T\mathbf{x}$\\
$(\mathbf{e}'_{4}+3\mathbf{e}'_{5}+4\mathbf{e}'_{6}+2\mathbf{r}_2)^T\mathbf{x}$ & $(\mathbf{e}'_{4}+4\mathbf{e}'_{5}+\mathbf{e}'_{6}+4\mathbf{r}_2)^T\mathbf{x}$ \\
\red $(\mathbf{r}_1+3\mathbf{r}_2+4\mathbf{r}_3)^T\mathbf{x}$ & \red $(\mathbf{r}_1+4\mathbf{r}_2+\mathbf{r}_3)^T\mathbf{x}$ \\
\blue $(\mathbf{e}'_{3}+3\mathbf{r}_4)^T\mathbf{x}$ & \blue $(\mathbf{e}'_{3}+4\mathbf{r}_4)^T\mathbf{x}$ \\
\blue $(\mathbf{e}'_{6}+3\mathbf{r}_5)^T\mathbf{x}$ & \blue $(\mathbf{e}'_{6}+4\mathbf{r}_5)^T\mathbf{x}$ \\
\blue $(\mathbf{r}_3+3\mathbf{r}_6)^T\mathbf{x}$ & \blue $(\mathbf{r}_3+4\mathbf{r}_6)^T\mathbf{x}$ \\
\end{tabular}
\caption{The responses sent by the servers when using an $(n,k,t)=(4,2,1)$ Staircase-PIR scheme.}
\label{tab:uniex}
\vspace{-0.3cm}
\end{table}

After receiving $\mathbf{q}_j$, each server $j$ projects the data $\mathbf{x}$ on $\mathbf{q}_j$ and sends the result back to the user. We illustrate how the user can retrieve the wanted file by achieving the PIR capacity $C(t,\mu)$ given in~\eqref{eq:acap} simultaneously for $\mu_1=4$, $\mu_2=3$ and $\mu_2=2$.

Suppose the user waits for $\mu_1=4$ servers. The user downloads the first $\alpha'/\alpha_1=2$ responses of each server corresponding to $(\mathcal{V}  \mathcal{M}_1)^T \mathbf{x}$ (first two rows in black in Table~\ref{tab:uniex}). Recall that $\mathcal{V}$ is a Vandermonde matrix, hence is invertible. The user multiplies the received responses by the inverse of $\mathcal{V}$ to decode $\mathcal{M}_1^T \mathbf{x}$ which contains ${\mathbf{e}'_j}^T\mathbf{x}$ for $j=1,\dots,6$, therefore retrieving the desired file. The rate of this PIR scheme is equal to $6/8=3/4 = C(1,4)$, because the user decodes the $6$ parts of the file by downloading $8$ responses. 

If the user waits for $\mu_2=3$ servers indexed by $L\subset [n]$, the user downloads the first $\alpha'/\alpha_2=3$ responses of each contacted server corresponding to $\left(\mathcal{V}_L \bbm \mathcal{M}_1 \ \mathcal{M}_2 \ebm\right)^T \mathbf{x}$ (in black and red), where $\mathcal{V}_L$ is the matrix formed by the rows of $\mathcal{V}$ indexed by $L \subset [n]$. Recall that $\mathcal{V}_L$ here is a $3\times 4$ Vandermonde matrix. The user can retrieve the file as follows. Since $\mathcal{M}_2$ has a $0$ as its last entry, $\left(\mathcal{V}_L \mathcal{M}_2\right)^T \mathbf{x}$ reduces to $\left(\mathcal{V}'_L \mathcal{M}_2\right)^T\mathbf{x}$, where $\mathcal{V}_L'$ is the $3\times 3$ invertible Vandermonde matrix formed of the first three columns of $\mathcal{V}_L$. The user can then decode $\mathcal{M}_2^T\mathbf{x}$ which consists of $\mathbf{r}_1^T\mathbf{x}$, $\mathbf{r}_2^T\mathbf{x}$ and $\mathbf{r}_3^T\mathbf{x}$. By subtracting $\mathbf{r}_i^T\mathbf{x}$ from $\left(\mathcal{V}_L \mathcal{M}_1\right)^T\mathbf{x}$, the user obtains $\left(\mathcal{V}_L'{\mathcal{M}_1'}\right)^T \mathbf{x}$ where $\mathcal{M}_L'$ is the matrix formed by the the first three columns of $\mathcal{M}_1$. By inverting $\mathcal{V}_L'$ the user can decode ${\mathcal{M}'_1}^T\mathbf{x}$ which contains the required file. The rate of this PIR scheme is equal to $6/9=2/3 = C(1,3)$, because the user decodes the $6$ parts of file by downloading $9$ responses. 

Following a similar procedure, the user can retrieve the file by downloading all the responses from any $2$ servers. The rate here is $C(1,2) = 1/2$ because the user downloads $12$ responses to decode the requested file. 

On a high level, privacy is guaranteed because each sub-query is padded with a different random vector.
\end{example}

\section{From secret sharing to PIR}\label{sec:ss}

The connection between secret sharing and PIR has been studied in the literature, e.g., \cite{WY05,beimel2002robust,DR18}.
Our Staircase-PIR construction was obtained using Staircase codes for communication efficient secret sharing \cite{BRIT18}. In this section, we explore more this connection between communication efficient secret sharing \cite{BR16,WW08,HLKBtrans,ZYSMH12,rawat2016centralized} and capacity achieving robust PIR schemes. 

\begin{definition}[Secret sharing scheme]
A secret sharing scheme is an encoding of a secret $\mathbf{s}$ into $n$ shares $\mathbf{w}_1, \dots, \mathbf{w}_n$ stored on $n$ servers, such that a user accessing any subset of $t$ or less shares obtains no information about $\mathbf{s}$, however by accessing any collection of $k$ or more shares the user can reconstruct the whole secret. Let $\texttt{S}$ denote the random variable representing the secret $\mathbf{s}$ and $\texttt{W}_i$ denote the random variable representing the share $\mathbf{w}_i$. A secret sharing scheme satisfies the following properties:

\noindent {\em 1) Perfect secrecy:} expressed as 
\begin{equation}\label{eq:secrecy1}
H(\texttt{S} \mid \texttt{W}_T)=H(\texttt{S}),\, \forall T\subset [n] \text{ s.t.}   \left\lvert T \right\rvert = t.
\end{equation}

\noindent {\em 2) MDS:} or reconstruction of the secret,
\be 
\label{eq:mds1}
H(\mrS \mid \texttt{W}_{A})=0,\, \forall A\subseteq [n] \text{ s.t.}   \left\lvert A\right\rvert=k,
\ee
{\Redit and the secret is of size $(k-t)$ units as implied by~\eqref{eq:secrecy1} and \eqref{eq:mds1} (see \cite[Proposition~1]{HLKBtrans}).}
\end{definition}
We refer to a secret sharing scheme as defined above as an $(n,k,t)$ secret sharing.
\begin{definition}[Communication efficient secret sharing] A communication efficient secret sharing is a secret sharing scheme that allows the user to reconstruct the secret by downloading a part of any $d$ shares, $k\leq d \leq n$. When accessing $d$ servers, the user needs only to download the optimal rate $d(k-t)/(d-t)$ units of information given in \cite{HLKBtrans,WW08}.
\end{definition}

Now we are ready to describe how to obtain a robust PIR scheme from a secret sharing (SS) scheme. We call this construction SS-PIR construction.

\noindent {\bf SS-PIR construction:} An $(n,k,t)$ robust PIR can be constructed using linear secret sharing as follows. Let $f_i$, the $i^{\text{th}}$ entry of the data $\mathbf{x}$, be the file of interest expressed as $f_i = x_i = \mathbf{e}_i^T \mathbf{x}$. To construct an $(n,k,t)$ robust PIR scheme, the user encodes the queries $\mathbf{q}_1,\dots,\mathbf{q}_n$ using an $(n,k,t)$ {\em linear} secret sharing scheme with $\mathbf{s}=\mathbf{e}_i$ and sends those queries to the $n$ servers. After receiving the query, each server $j$ projects the data $\mathbf{x}$ on $\mathbf{q}_j$ and sends $\mathbf{q}_j^T\mathbf{x}$ to the user.

\begin{proposition}\label{prop:pirsec}
An information retrieval scheme obtained from the SS-PIR construction is a robust PIR scheme, i.e., guarantees privacy and robustness and achieves asymptotic capacity equal to
\begin{align*}
C(t,k)=1-\frac{t}{k}.
\end{align*}
\end{proposition}

We give the proof of Proposition~\ref{prop:pirsec} in the Appendix. We generalize Proposition~\ref{prop:pirsec} to show that a PIR scheme constructed using a communication efficient secret sharing is universally robust.

\begin{proposition}\label{prop:urpircess}
An information retrieval scheme constructed using an $(n,k,t)$ linear communication efficient secret sharing scheme is a universally robust PIR that achieves PIR capacity
\begin{align*}
C(t,\mu)=1-\frac{t}{\mu},
\end{align*}
simultaneously for all $k\leq \mu \leq n$.
\end{proposition}
The proof of Proposition~\ref{prop:urpircess} can also be found in the Appendix. The implication of this Proposition is that any communication efficient secret sharing including Staircase codes can be used to obtain universally robust PIR through SS-PIR construction.

\section{Conclusion}\label{sec:conc}

We study robust private information retrieval (PIR). A user wants retrieve a file by querying $n$ servers without revealing the identity of the required file to the servers. We assume the servers can collude and consider the setting in which the servers might be  stragglers, i.e., slow or unresponsive. We introduce Staircase-PIR, a universally robust PIR scheme that allows the user to successfully retrieve the file by waiting only for the non straggler servers. This scheme achieves the PIR capacity simultaneously for any number of stragglers up to a given threshold. Moreover, we give a general construction to obtain universally robust PIR from communication efficient secret sharing.

\bibliographystyle{ieeetr}
\bibliography{IEEEabrv,DSS}

\appendix

\subsection{Proof of Proposition~\ref{prop:pirsec}}  We show that an information retrieval scheme constructed using linear secret sharing scheme is a robust PIR scheme, i.e., guarantees privacy and robustness, and achieves asymptotic PIR capacity.

\noindent{\em Privacy:} Any subset of $t$ or less servers obtain no information about the identity of the file of interest. Each server observes a query encoded using an $(n,k,t)$ secret sharing. From the secrecy constraint of secret sharing~\eqref{eq:secrecy1}, any $t$ or less servers obtain no information about $\mathbf{e}_i$, which represents the identity of the file.

\noindent{\em Robustness:} After receiving the responses $\mathbf{q}_j^T\mathbf{x}$ from any $k$ servers, $j\in[n]$, the user can retrieve $f_i$. From the MDS property of the secret sharing~\eqref{eq:mds1}, the user can decode $\mathbf{e}_i$ from any $k$ queries $\mathbf{q}_j$, $j\in [n]$. By linearity of the secret sharing scheme, after receiving $k$ responses $\mathbf{q}_j^T\mathbf{x}$ from the servers, the user is able to decode $f_i = \mathbf{e}_i^T\mathbf{x}$. In other words, since all the queries are multiplied by the same vector $\mathbf{x}$, being able to decode the secret $\mathbf{s}=\mathbf{e}_i$ from the queries implies the ability of decoding the file $\mathbf{e}_i^T\mathbf{x}$ from the responses, c.f. Section~\ref{sec:uniex}.

\noindent{\em Optimality:} We show that this PIR scheme is optimal, i.e., achieves $C(t,k)$ given in~\eqref{eq:acap}. To retrieve the file, the user has to download $k$ responses, i.e., $k$ units of information. Secret sharing assumes that the size of the retrieved file is $k-t$ units of information\footnote{The scheme can be scaled so that the file is of size $1$ unit of information, each query becomes of size $1/(k-t)$.}. Therefore, the rate of this PIR scheme is ${(k-t)}/{k}$ achieving $C(t,k)$.

\subsection{Proof of Proposition~\ref{prop:urpircess}}
We show that an information retrieval scheme encoding the queries using a linear communication efficient secret sharing scheme is an $(n,k,t)$ UR-PIR. Note that privacy and robustness to any number of unresponsive servers are guaranteed by the properties of secret sharing and by the ability of decoding the secret by accessing any $d$ shares. The additional property that we comment on is the optimality of retrieving the file when any number of servers $\mu$, $k\leq \mu\leq n$, are stragglers. When a user receives responses from $\mu$ servers it only needs to download $\mu(k-t)/(\mu-t)$ units of information to retrieve the file of size $k-t$. Therefore, the rate of this scheme is equal to $(\mu-t)/\mu$ achieving~\eqref{eq:acap} with equality.

\end{document}